\documentclass[12pt]{article}
\usepackage{amsmath}
\usepackage{latexsym}
\usepackage{amssymb}
\usepackage{fullpage}
\usepackage{lscape}
\usepackage{tikz}
\usepackage{setspace}
 \usepackage{colortbl}
\usepackage{epsfig}
\usepackage{graphics}

\usepackage[a4paper,bindingoffset=0.2in,%
            left=0.6in,right=0.6in,top=1.15in,bottom=1.15in,%
            footskip=.25in]{geometry}
\usepackage{blindtext}

\newcommand{\qed}{\hfill \ensuremath{\Box}}
\def\dimo{\noindent\mbox{\sc proof: }}
\def\eproof{\rm\hspace*{\fill}$\Box$\vspace{10pt}}
\newtheorem{defin}{\bf Definition}[section]
\newtheorem{theo}[defin]{Theorem}
\newtheorem{lem}[defin]{Lemma}
\newtheorem{corol}[defin]{Corollary}
\newtheorem{prop}[defin]{Proposition}

\newtheorem{oss}[defin]{Remark}
\newtheorem{ex}[defin]{Example}

\def\cV{{\cal V}}

\def\cW{{\cal W}}
\def\cP{\mbox{\boldmath ${\cal P}$}}
\def\cS{{\cal S}}

\def\cU{{\cal U}}

\def\cI{{\cal I}}

\def\meglio {{\quad {\underset {\sim}  \succ}}}

\def\cccell{\mbox{\boldmath $\ell$}}
\def\ccck{\mbox{\boldmath $k$}}

\def\cccx{\mbox{\boldmath $x$}}

\def\cccy{\mbox{\boldmath $y$}}
\def\cccz{\mbox{\boldmath $z$}}
\def\cV{\mbox{\boldmath ${\cal V}$}}
\def\ccD{\mbox{\boldmath ${\cal D}$}}
 \def\cVc{\mbox{\boldmath $\widehat{\cal V}$}}
\def\cSigma{\mbox{\boldmath ${\Sigma}$}}

\def\bDelta{\mbox{\boldmath ${\Delta}$}}
\def\include{\mbox{\boldmath $\ge$}}
\def\meglio {{ {\underset {\sim}  \succ}}}
\def\cccfi{\mbox{\boldmath $\Phi$}}

\DeclareRobustCommand{\rchi}{{\mathpalette\irchi\relax}}
\newcommand{\irchi}[2]{\raisebox{\depth}{$#1\chi$}} 
 
\begin{document}

 \newpage

\title{ {\sc The character of
 non-manipulable collective choices between two alternatives
 }}

\author{Achille Basile\thanks{Corresponding author, Dipartimento di Scienze Economiche e Statistiche,  Universit\`a Federico II,
80126 Napoli, Italy, E-mail: basile@unina.it;  financial support from the Project 000004 Star Linea 1 - 2020 is gratefully acknowledged.},\,
{K. P. S. Bhaskara Rao\thanks{Department of Computer Information Systems, Indiana 
University Northwest, Gary, IN 46408, E-mail: bkoppart@iun.edu}},\,
{Surekha Rao\thanks{School of Business and Economics, Indiana University Northwest,
Gary, IN 46408, E-mail: skrao@iun.edu}} ,\,
}

 

\maketitle

\thispagestyle{empty}

 \begin{abstract}

  {\color{black} We consider classes of non-manipulable social choice functions with range of cardinality at most two within a set of at least two alternatives. We provide  the functional form for each of the classes we consider. 
 This functional form   
  is a characterization that 
  explicitly describes how a social choice function of that particular class selects the collective choice corresponding to a profile. 
 We provide a unified formulation of these characterizations using the new concept of \lq\lq character\rq\rq. The choice of the character, depending on the class of social choice functions, gives the functional form of all social choice functions of the class.}

\end{abstract}

JEL Code: D71


{\it{Keywords: 
social choice functions, weak group strategy-proofness,  anonymity,  functional form, character function,
  preferences, restricted domain. 
}}

\bigskip

 \newpage
 
  \setcounter{page}{1}

\section{Introduction} 

{\color{black}
 
 The characterization of 
 classes of social choice functions is a central theme in Social Choice Theory. 
 There are characterizations that describe in an explicit way the rule that determines the collective choice once a profile of preferences is chosen.

 Such kind of characterizations are referred to as {\it functional form} characterizations by Barber\`a et al.  \cite {BBM}, whereas  Hagiwara and Yamamura \cite{HY} call them {\it closed} characterizations.

In the theory of non-manipulable social choice functions there are classical results which really are functional form characterizations. The celebrated Gibbard-Satterthwaite theorem is an example. When the collective choice is taken between two alternatives and anonymity of agents is required, the quota majority method is another example. A classical reference is  Moulin \cite{M}.
 
This paper focuses on the functional form of non-manipulable collective choice between two alternatives. We consider  a model for  social choice in which the realizable alternatives are two, indifference is permitted, and the collective decision is based upon a preference profile of agents who also consider  other alternatives.  

As compared to Barber\`a et al. \cite {BBM}, and Hagiwara and Yamamura \cite{HY}, we do not restrict the set of agents to be finite.  As such non-manipulability is based on groups of agents rather than on a single agent.

 We deal mainly with {\it weakly group strategy-proof} social choice functions, according to Definition 2 in  Barber\`a et al. \cite{BBM}. 
 It is well known
   that  this class of functions does not coincide in general with the class of  individually strategy-proof social choice functions.\footnote{Even if one assumes that there are only finitely many agents, some assumptions are necessary in order to ensure that individual strategy-proofness is equivalent to weak group strategy-proofness.} 
However, since we only deal with social choice functions with range of cardinality two,  weak group strategy-proofness is the appropriate non-manipulability notion for the case of  infinite agents (see
  \cite[Corollary 1]{BBM10} and \cite[Theorem 3.3. (ii)]{KPS-EL}). Moreover, our results extend to individual strategy-proofness for finitely many agents.

\medskip
 {\bf Contributions of the paper.}
 Although Barber\`a et al. \cite{BBM}  characterize non-manipulability, they do not provide the functional form of non-manipulable social choice functions.  Hagiwara and Yamamura \cite{HY} present as a closed characterization their main result Theorem 2. Neither papers considers anonymity.

 The two of the three significant contributions of our paper are: we provide the functional form of
 
\begin{itemize}
\item [$(1)$] all non-manipulable social choice functions;

\item [$(2)$] all non-manipulable, anonymous social choice functions.
\end{itemize}

The third notable contribution is the notion of {\it character},  which we use to introduce the {\it canonical} functional form of social choice functions  with range of cardinality two within a set of at least two alternatives. The functional forms of the classes $(1)$ and $(2)$ can be unified, being both canonical under an appropriate choice of a character. The unifying nature of the canonical functional form is demonstrated by discussing several  classes of social choice functions.

With these three results,
our paper completes the work initiated by Barber\`a et al. \cite{BBM}, 
and further developed  by Hagiwara and Yamamura \cite{HY}. In addition, it significantly generalizes
 results of Basile et al. \cite{KPS-GEB}, for the model with only two alternatives.

 \medskip
 {\bf Relevance of the setting.}  
  Analyzing the non-manipulable collective choice between two alternatives $a$ and $b$ (candidates, public projects, the introduction of a new law versus the status quo, ...) is different if agents are asked to report their preference just between $a$ and $b$,
   or if they are asked, as we assume in our more general setting, for preferences over a larger set
 of alternatives that includes $a$ and  $b$.
The fact that the collectivity \lq\lq implements\rq\rq\, either $a$ or $b$, may induce to think that the social choice is independent of the remaining alternatives (principle of independence of irrelevant alternatives). This is not always the case, even if strategic reporting is ruled out by the  social choice functions adopted.
The following example  demonstrates this.

\begin{ex}\label{Esempio 11} {\sl There are only two agents $v_1$, and $v_2$ and three alternatives $a, b,$ and  $c$. The social choice function  we consider for the collective decision of implementing either $a$ or $b$ is as described next.
Corresponding to a profile of preferences over the three alternatives, the collective choice is $b$ if and only if either one of the following circumstances is true:

-- agent $v_1$ prefers $b$ to $a$;

-- agent $v_1$ is indifferent between $a$ and $b$ but prefers $c$ to both $a, b$;

In all the other cases the rule selects $a$.

Neither agent may gain by lying, i.e. this scf  is non-manipulable.  

 Now, consider the following two profiles
$P$ and $Q$. In profile $P$: agent $v_1$ is indifferent between $a$ and $b$ that are both worse than $c$. 
 In profile $Q$: 
 agent $v_1$ is indifferent between $a$ and $b$, but  both alternatives are strictly preferred to  $c$; agent $v_2$ reports the same preference as in profile $P$. 
 
 Even if both profiles are indistinguishable between $a$ and $b$, the collective choice described above selects $b$ for $P$ and $a$ for $Q$.}\eproof
  \end{ex}
 Hence a non-manipulable social choice function does not guarantee the independence of the collective choice from the irrelevant alternatives. In particular, this tells us that investigating the setting of the present paper is not obvious from Basile et al. \cite{KPS-GEB}. 

 \bigskip
On the other hand, considering our model,  is not merely an intellectual exercise.  Some arguments supporting its  relevance   are presented by  Barber\`a et al. \cite{BBM} and by   Hagiwara and Yamamura \cite{HY}. We like to add that it can be  also seen as  a model for  balancing  simplicity  and full attention to agents' necessities. An intuition that supports this idea  is the following, simple, real-life situation.
Suppose that a community has to decide about a public project like building an educational institution.   \lq\lq Educational Institutions\rq\rq\, differ by many dimensions.  They differ by students they target (primary, secondary, a college, ...), or by learning environments provided for teaching e.g;  online, in person, blended, residential,  vocational or other format. Of course they differ by location, and so on.

 So, many potential alternatives are on the table. 
Naturally, every citizen has his/her opinion about the various options.  With respect to some, they may be indifferent.
Even if the potential
alternatives are many,  for several reasons, the designer of the public choice mechanism may limit the outcome of the collective decision only to either build solution $a$ or build solution $b$. 
Nonetheless, the implementation of a social choice function  that asks for preferences on the entire spectrum of possibilities rather than only on $\{a,b\}$, allows us to take into account the preferences of the agents also regarding  the other alternatives.
By doing so, there is a kind of balance between simplicity (either $a$ or $b$) and respect for the needs of citizens (the input for the social choice is a profile of preferences over all potential alternatives).
There is only one possible drawback of this idea: if agents are asked preferences over the entire set of potential alternatives, but only either $a$ or $b$ is realized, then it looks like  agents have rights that the social planner don't care. If we think carefully, we see that this is not the case, as shown in  Example \ref{Esempio 11} according to which the collective choice is not independent of the full spectrum of alternatives.
We believe that here clearly emerges  {\it \lq\lq the choice to restrict the range} (of the social choice function) {\it as a possible tool for the mechanism designer\rq\rq}, as emphasized by 
 Barber\`a et al. \cite{BBM} in their final remarks.

Together with the previous  argument (of philosophical flavor, say) in favor of the  study of our setting, there are also theoretical arguments. For example,  Barber\`a et al. in a different paper,  \cite{BBM12}, show that {\it a non-manipulable social choice function, over the domain made of all profiles consisting of single-dipped preferences relative to some linear order, necessarily has range of cardinality at most two }. 

}

\bigskip
The plan of the paper is as follows. After presenting in Section \ref{sezione2}  the social choice model we adopt, Section \ref{tre} is devoted to obtain the representation of general non-manipulable social choice functions with range of cardinality at most two. We also introduce the notions of character and canonical functional forms. In Section \ref{ancase} we analyze the anonymous case. 
Section \ref{unifying}, by  considering several classes of social choice functions, demonstrates how the character function approach can be  systematically applied to them. This also shows the role of the character in  unifying  the functional form characterizations of these classes.

The paper ends with Conclusions and an Appendix with some  proofs.

\section{The social choice model}\label{sezione2}
{\color{black}Let $V$ be a  set representing  a collectivity whose members we refer to as {\it agents}.
Let $\cW(A)$ be the set of all  complete, transitive binary relations over an arbitrary set $A$ of alternatives.  We refer to such relations as {\it preferences}.
With reference to   a preference  $W\in\cW(A) $,  we adopt standard  notations: 
$ x\quad {\underset {\sim}  \succ}_{W} y$  stands for $(x,y)\in W$, the notation $ x\, \succ_W \,y$ stands for $\big[(x,y)\in W$ and $(y,x)\notin W\big]$, and the notation $ x\, \sim_W \,y$ stands for $\big[(x,y)\in W$ and $(y,x)\in W\big]$.

Functions $P$ from $V$ to $\cW(A)$ are named {\it  preference profiles}. The class of all  preference profiles 
is denoted by $\cW(A)^V$. A {\it partial profile of preferences} is a function 
$\pi$ from a subset of $V$ to $\cW(A)$. A partial profile is therefore $\pi=(\pi_v)_{v\in T}$ where $T$ is a subset of $V$ and $\pi_v\in \cW(A)$ for every $v \in T$.
  
A particular partial profile is that which involves the empty set as domain; we speak of this as the {\it empty profile}. }

For a profile $P=(P_v)_{v\in V}$ we shall also use the notation  $P=[P_T, P_{T^c}]$   if $T$ is a subset of $V$, the set $T^c$ is its complement, and $P_T$, $P_{T^c}$ are the obvious restrictions $P_T=(P_v)_{v\in T}$, $P_{T^c}=(P_v)_{v\notin T}$ of $P$. Extending this notation to arbitrary partitions of $V$ or to partial profiles is straightforward.

\bigskip
Let $\cP$ be a subset of $\cW(A)^V$. We refer to $\cP$ as  the class of feasible profiles.  Possible assumptions on $\cP$ are:\begin{itemize}

\item [] {\it universal domain}:
$\cP=\cW(A)^V$, 
\item [] {\it cartesian restricted domain}: $\cP=\times_{v\in V}\cW_v$ with $\O\neq\cW_v\subseteq \cW(A)$, 

\item []  {\it quasi--cartesian restricted domain}:   the set $\cP$  has the property $$P, Q\in \cP, T\subseteq V\Rightarrow [P_{T}, Q_{T^c}]\in \cP.$$
\end{itemize}

Trivially, cartesian implies quasi--cartesian. For a finite set of agents, the converse is also true.  In the case of infinitely many agents,
quasi--cartesian  is  a weaker assumption than cartesian.\footnote{ For example, let $V$ be the set of natural numbers and $W_0,W_1$ two distinct preferences. The domain $\cP$ consisting of all profiles $P$ that can be written as 
$ P_v= \left\{ 
\begin{array} {ll}
W_1 , & \mbox { if } v\in F  \\

W_0 , & \mbox { otherwise }  \\
\end{array}
\right.$ for some finite subset $F$ of $V$, is quasi-cartesian but not cartesian.
} 

Note that the possibility of dealing with strict preferences (i.e. with the elements of the subset $\cS(A)$ of $\cW(A)$ made of preferences that are also antisymmetric) is not excluded in the case of cartesian restricted domain (or quasi--cartesian). The case $\cP=\cS(A)^V$, will be referred to as {\it strict universal domain}.

\bigskip
 Throughout the sequel,  $\phi: \cP\to A$ stands for a social choice function (scf, for short). If the range of $\phi$ has cardinality at most two, we say that $\phi$ is a two-valued scf. 

\subsection{Non-manipulability.}\label{21}
{\color{black}
Our setting coincides with that of Barber\`a et al. \cite{BBM} except for the fact that the sets $V$, of agents, and $A$, of alternatives,  need not to be  finite.

Dealing with an arbitrary set $V$, the notion of non-manipulability we adopt is based on coalitions of agents rather than on a single agent. A \lq\lq coalition of agents\rq\rq\, stands for a \lq\lq non-empty group of agents\rq\rq.
A desirable property of a scf  is that there are no incentives for the  agents  to coalesce forming a group  
that, with false reporting, can manipulate the social outcome for the advantage of its own members. Groups can organize manipulation in several ways. Because of this,  Barber\`a et al. \cite{BBM} present different notions of group manipulations. We adopt the following notion of  group manipulation.

\begin{defin}\label{manipulation}
Let $\phi$ be scf.
 We say that a coalition $D$ can strongly manipulate a profile $P\in \cP$   under $\phi$  if there is another profile $Q\in\cP$ such that 

\begin{itemize}
\item every agent $v$ in $D^c$ has the same preference in both $P$ and $Q$, \, i.e. $P_v=Q_v$; 

\item every agent $v$ in $D$ prefers $\phi(Q)$ to $\phi(P)$ according to $P_v$, \, i.e. $\phi(Q)\succ_{P_v} \, \phi(P)$.
\end{itemize}
\end{defin}

The impossibility of the above form of manipulation leads to the notion of weak group strategy-proofness.

\begin{defin}\label{wGSP}
 We say that a scf $\phi$ is  {\bf  weakly group strategy-proof}  if no coalition of agents can strongly manipulate any feasible profile under $\phi$.  Moreover, we shorten the expression \lq\lq weakly group strategy-proof\rq\rq\, as wGSP.
\end{defin}

When $\phi$ is wGSP, we also say sometimes that $\phi$ is non-manipulable.
Evidently, the above weak group strategy-proofness of a scf $\phi$ is the same as the validity of the following implication for $\phi$:
$$
\big[ D \mbox{ is a coalition, } P, Q\in \cP \mbox{ are identical on } D^c   \big] \Rightarrow \big[ \exists v\in D: \,\phi(P) \meglio_{P_v} \phi(Q)    \big].
$$

We have adopted the terminology of  Barber\`a et al.  \cite[Definition 2]{BBM}. However, note that the term \lq\lq coalitional strategy-proofness\rq\rq\, is sometime used for the same concept (see \cite{LBZ} and \cite{KPS-JME}, for example).
}

A further  property equivalent to wGSP is the following

 \def\gam{\left\{ \begin{array} {ll}
\phi(P)&\meglio_{P_v} \phi(Q)  \\
P_v&\neq \,Q_v
\end{array}
\right.}

$$P, Q\in \cP,\,\phi(P)\neq\phi(Q) \implies \exists v\in V \mbox{ such that } \gam$$
introduced in \cite{KPS-EL} under the name {\it almost preference reversal} (APR, for short).

\bigskip
In the previous Definition \ref{wGSP}, if we replace coalitions with singletons, the corresponding notion is known  as
{\it individual strategy-proofness}. 
\begin{defin}
A scf $\phi$ is
{\bf individually strategy-proof} (ISP, for short), if, according to $P_v$, the alternative \,\, $ \phi(P_v, P_{-v}) $ is at least as good as $ \phi(Q_v, P_{-v})$, for every agent $v$, for every profile $P\in \cP$, and for every preference $Q_v$ such that $(Q_v, P_{-v})\in\cP$.
\end{defin}

{\color{black}
It is well known that wGSP and ISP are not equivalent notions in general. Even if we restrict the setting by assuming that $V$ is finite, further assumptions are necessary in order to ensure that ISP implies wGSP. Le Breton and Zaporozhets \cite {LBZ} and Barber\`a et al. \cite{BBM10} investigates this in depth.} However, some plain cases of equivalence of the two notions can be  recalled without resorting to technicalities. 

Suppose $V$ is finite: 

-- ISP implies wGSP when $\cP$ is universal (as it can be verified with a direct proof);

-- ISP implies wGSP when $\cP$ is a cartesian restricted domain but the scf has  range of cardinality at most three (\cite[Corollary 1]{BBM10} and \cite[Theorem 3.3 {\it (ii)}]{KPS-EL}).
{\color{black}
\bigskip
\begin{oss}\label{ispwGSP}
{\sl  We concentrate our attention on  weak group strategy-proofness. This property is more stringent  than individual strategy-proofness but it allows us to skip the assumption of the finiteness of $V$ in our results.
 The statements we have just recalled above tell us that when $V$ is finite our results extend to the case of individual strategy-proofness.  In a sense this is a point of view opposite to Hagiwara and Yamamura \cite{HY}. They first obtain  the closed  characterization of ISP scfs, and, on the basis of \cite[Corollary 1]{BBM10}, present  it as the closed characterization of wGSP scfs (which they name group strategy-proof scf).}
\end{oss}
}

\section{The functional form of all two-valued weakly group \\ strategy-proof social choice functions}\label{tre}
In this section we obtain the canonical representation of a   two-valued wGSP scf.  Without loss of generality,
we adopt the following approach. {\it For every set $\{a, b\}\subseteq A$ consisting of two distinct alternatives 
we determine the functional form of the wGSP social choice functions with range contained in  $\{a, b\}$.} 

\medskip
With this in mind,   we adopt the following notation: by $D(x, P)$    we denote the subset of $V$ consisting of agents that, under the profile $P$, prefer the alternative $x\in \{a, b\}$ to the other remaining alternative in $\{a, b\}$; by $I(P)$  we denote the subset of $V$ consisting of agents that, under $P$ are indifferent between $a$ and $b$. We also set $ \overset\sim D(x, P)=D(x, P)\cup I(P)$.

\subsection{Monotonicity}\label{31}
{\color{black}The identification of the functional form of   two-valued wGSP scfs goes through a  characterization of non-manipulability  by means of a
 {\it monotonicity}  \`a la Maskin: suppose that the society selects $a$ when profile $Q$ prevails, then it is natural to ask that $a$
is again selected under a new profile $P$ that shows no less support for the alternative $a$. A relevant issue is the precise meaning that needs to be given to the  verbal expression 

$$(\maltese) \mbox{  {\it \lq\lq the profile $P$ supports the alternative $a$ at least as the profile $Q$ does\rq\rq}.}$$

In the following we discuss this, demonstrating the value this paper add to Basile et al. \cite{KPS-GEB}.

{\bf Example 1.1 (continuation).} {\sl Comparing the model adopted in \cite{KPS-GEB} to the one adopted here, we note that here we do not assume that $A=\{a, b\}$. 

In  \cite{KPS-GEB}
by the sentence $(\maltese)$  
we exactly meant that,
\begin{itemize}
\item[$(\maltese_{bi})$] changing the profile of preferences of the society from $Q$ to $P$, agents favoring  $a$ against $b$
remain so, and agents that were indifferent between $a$ and $b$, and have not moved to prefer $a$, remain indifferent between $a$ and $b$.
\end{itemize}
On the basis of this, the monotonicity  (\`a la Maskin) characterizes non-manipulability \cite[Theorem 2.,5]{KPS-GEB}.
However, in Example \ref{Esempio 11}  we see that even if the support for the alternative $a$ shown by $P$ is not less than the one shown by $Q$, the change of profile from $Q$ to $P$ changes the collective choice from $a$ to $b$. So the scf of that example is not monotone.
In other words,
in the  setting of the present paper Theorem 2.5  of \cite{KPS-GEB} {\it verbatim} does not hold true. We need  to enlarge the class of monotone scfs in order to keep the equivalence between monotonicity and non-manipulability.  
To these goal the present section is devoted.} \eproof

\medskip
We propose a more stringent interpretation of the sentence $(\maltese)$. In contrast to $(\maltese_{bi})$
in  our more general setting we change the meaning as follows:

\begin{itemize}
\item[$(\maltese_{two})$] changing the profile of preferences of the society from $Q$ to $P$, agents  in favor of $a$ against $b$ remain such, and agents that were indifferent between $a$ and $b$, and have not moved to prefer $a$, {\bf do not change their opinion about the entire set of alternatives}.
\end{itemize}

Based on this, a   weaker, yet straightforward, notion of monotonicity can be introduced that characterizes the non-manipulability (Theorem \ref{49}, {\it infra}).

}

The following definition formalizes the meaning of $(\maltese_{two})$.

{\color{black}\begin{defin}\label{AE11}
Given two profiles $P$ and $Q$, we say that $P$ supports the alternative $a$ at least as  $Q$ does, and write $P\overset a\ge Q$, if
$$D(a, Q)\subseteq D(a, P), \mbox { and } \,\, v\in  I(Q)\setminus D(a, P) \mbox{ implies that }P_v=Q_v.$$
Replacing $b$ to $a$ in the two conditions above, we define the relation $P\overset b\ge Q$.
\end{defin}

 The relation $P\overset b\ge Q$  can be interpreted similarly: {\it \lq\lq the profile $P$ supports the alternative $b$ at least as the profile $Q$ does\rq\rq\,}. The fact that in our exposition we privilege the reference to $a$ in the definition of monotonicity,  does not affect the generality of our results. 
 
 It is obvious that: $P\overset a\ge Q \Leftrightarrow Q\overset b\ge P$. }

The next definition introduces the mentioned extension of the monotonicity property used in \cite{KPS-GEB}. To emphasize that the new definition is weaker than  the monotonicity property in \cite{KPS-GEB}, we refer to it as {\it almost monotonicity}.

\begin{defin}\label{monotonicity}
Let $V, \cP, A$ be arbitrary and $a, b$ two distinct elements of $A$. We say that a
  scf $\phi:\cP\to \{a, b\}$  is  almost monotone if $$\big[P,Q\in\cP,\,P\overset a\ge Q, \quad \phi(Q)=a\big]\Rightarrow \phi(P)=a.$$
\end{defin}
The relation of almost monotonicity with non-manipulability is the focus of the next theorem.
\begin{theo}\label{49}
Let $a, b$ be two distinct elements of $A$. Consider a
  scf $\phi:\cP\to \{a, b\}$. Then, $\phi$    is  wGSP   if it is almost monotone. Conversely, on a quasi--cartesian restricted domain, $\phi$    is   almost monotone if it is wGSP.
\end{theo}
\dimo
We show that almost monotonicity implies almost preference reversal.
Suppose $\phi$ is not APR.  By definition this means that we find feasible profiles $P,Q$ such that  $\phi(P)\neq\phi(Q)$ and moreover with, for every $v\in V$, either $\phi(Q) \succ_{P_v} \phi(P)$ is true or $P_v=Q_v$ is true. 

If $\phi(P)=a$ we get $P\overset b\ge Q$. If $\phi(P)=b$ we get $P\overset a\ge Q$. This is immediate. In both cases almost monotonicity is violated.

Now, under the assumption that the scf is defined on a quasi--cartesian restricted domain, we prove the converse. So, let us assume that $\phi$ is wGSP and show, by contradiction, that it must be almost monotone. 

Suppose almost monotonicity is violated. Then we have two profiles $P,Q\in\cP$ such that $P\overset a\ge Q$, $\phi(Q)=a$, and $\phi(P)=b$.
Let us consider the following partition $\{V_0, V_1, V_2\}$ of $V$:

$V_0=\{v\in V: P_v=Q_v\},$ 

$V_1=\{v\in V: P_v\neq Q_v \}\cap D(b, Q), $

$V_2=\{v\in V: P_v\neq Q_v \}\cap \overset\sim D(a, Q).$

\bigskip
{\color{black}Since we have the hypothesis that $\phi$ is wGSP and this is the same as  APR, we have an agent $v$ (at least one exists because of APR)
for which one has both $P_v\neq Q_v$ and $b\,\,\meglio_{P_v} \!\!a$. In other words we have
\quad $\O\neq \overset\sim D(b, P)\cap \{v\in V: P_v\neq Q_v \} .$
Observe that the agents belonging to
$\overset\sim D(b, P)\cap \{v\in V: P_v\neq Q_v \} $ necessarily prefer $b$ to $a$ under $Q$. This is a consequence of  $P\overset a\ge Q$, i.e. of the validity of the two conditions $D(a,Q)\subseteq D(a, P)$ and $v\in I(Q)\setminus D(a, P)\Rightarrow P_v=Q_v$. So, we can write}
$$\O\neq \overset\sim D(b, P)\cap \{v\in V: P_v\neq Q_v \} \subseteq V_1.$$

Let us consider the profile $P'=[P_{V_0}, Q_{V_1}, P_{V_2}]\in \cP$. Since $\phi$ is wGSP, then $\phi(P')=b$. Otherwise the coalition $V_1$ manipulates $P'$ by means of $P$.

If the set $V_2$ is empty, then $P'= Q$ and we contradict $\phi(Q)=a$. So, it must be the case that $V_2$ is nonempty. However, we again contradict $\phi(Q)=a$. Otherwise, having $\phi(Q)=a$, the coalition $V_2$ manipulates $P'$ by means of $Q$ ({\color{black}observe that, again by the two conditions $D(a,Q)\subseteq D(a, P)$ and $v\in I(Q)\setminus D(a, P)\Rightarrow P_v=Q_v$, every agent in the coalition $V_2$ prefers $a$ to $b$ under $P$, i.e.  $V_2\subseteq D(a,P)$).}
\eproof

\begin{ex} \sl In order to prove the implication wGSP $\Rightarrow$ almost monotonicity,  we cannot dispense with the assumption on $\cP$ of being quasi--cartesian. Indeed, let us consider $V=\{v_1, v_2\}, A=\{a, b, c, d\},$ and the domain $\cP$ consisting of only two feasible profiles $P, Q$. 
Assume  $P$ is given by: $P_{v_1}= a\succ b\succ c\succ d$, and $P_{v_2}= b\succ a\succ d\succ c$. Assume  $Q$ is: $Q_{v_1}= c\succ d\succ a\succ b$, and $Q_{v_2}= d\succ c\succ b\succ a$. Define $\phi(P)=a, \phi(Q)=b$. The scf $\phi$ is APR hence wGSP but it is not almost monotone.\eproof\end{ex}

{\color{black}\begin{oss}\label{oss35}
\sl
With reference to scfs whose range consists of alternatives named $x,y$, Barber\`a et al (2012)  introduce the notion of essentially xy-monotonic   scf \cite[Definition 5]{BBM} and the notion of essentially xy-based scf \cite[Definition 7]{BBM}.
Considering that in the present paper we refer to $\{a, b\}$- valued functions, we observe, see \cite[Remark 2.12]{KPS-JME}, that our notion of almost monotonicity is equivalent to that of being simultaneously essentially ab-based  and essentially ab-monotonic.\end{oss}}

%
%

\subsection{The character function}
Before introducing all the \lq\lq ingredients\rq\rq\, needed for establishing the canonical representation of a two-valued non-manipulable scf, we recall some notions about general partially ordered sets.

\bigskip
{\color{black}We recall that a partially ordered set (poset, briefly) is a set in which  a binary relation is defined which is  reflexive, transitive, and antisymmetric. The notation we adopt for a poset is $L$. When we want to emphasize that on the set $L$, the underlying set that can be made partially ordered with the adoption of different binary relations,  we are considering a specific relation denoted by $\ge$, then we write $(L, \ge)$ to denote a poset. As  is customary, for $x,y\in L$, the validity of $x\ge y$  is also denoted by $y\le x$.}

 \begin{defin}\label{SOC}
 Let $(L, \ge)$ be a partially ordered set. 
 
  A subset $C$ of $L$ is {\bf super order closed}  if  
 $$ x\in L, f\in C, x\ge f \Rightarrow x\in C.$$ A subset $M$ of $L$ is an {\bf antichain} when
 $$x, x'\in M,\,\,x\neq x'\Rightarrow x'\ge x \mbox{ is false. }$$
 \end{defin} 
 
 When $C$ is a super order closed subset of a poset we  write $C$ is a SOC subset or $C$ is SOC, in short.
 
 Of course the empty set is an antichain and it is SOC. 

If $X$ is a subset of $L$, its subset $m(X)$, defined as  $m(X)=$ $\{x\in X: x\ge y, y\in X\Rightarrow y=x\}$, denotes the set of the minimal elements of $X$. Of course  $m(X)$ is an antichain.

Note that for a finite poset $L$ it is possible to prove the following (see the Appendix for the proof).

\begin{prop}\label{finito}  The antichains of a finite poset $L$ are in  one--to--one correspondence with the SOC subsets of $L$.\end{prop}

\begin{defin}\label{carattere}
 A character on the set $\cP$ of the feasible profiles is a function $\rchi$ from $\cP$ to the underlying set  $L$ of a poset $(L,\ge)$.
\end{defin}

By using a fixed character $\rchi$,  one can introduce two-valued scfs in a straightforward way. For each subset $C$ of the image set $\rchi(\cP)$, we define a scf.

\begin{defin}\label{forma canonica}
Let $a$ and $b$ be two distinct elements of $A$, and  $\rchi$  be a character. For every $C\subseteq \rchi(\cP)$, by setting 
$$\phi_{(\rchi, {\footnotesize C})}\,(P)=a\overset{\rm def}\Longleftrightarrow \rchi(P)\in C,$$
 we define a  function $\phi_{(\rchi, {\footnotesize C})}:\cP\to \{a, b\}$. 
The scf $\phi_{(\rchi, {\footnotesize C})}$ will be called the canonical scf.\end{defin}

When no ambiguity  occurs about the character, we shorten $\phi_{(\rchi, {\footnotesize C})}$ as $\phi_C$.

\bigskip
Now we introduce a notion of monotonicity of $\{a, b\}$-valued scfs with respect to a character. As in the previous notions of monotonicity we can equivalently choose to refer to $a$ or to $b$.
\begin{defin}
Let $\rchi$ be a character on the set $\cP$ of the feasible profiles.
A  scf  $\phi: \cP\to \{a, b\}\subseteq A$ is said to be
$(\rchi, a)$-monotone when

$$\big[P,Q\in \cP,\,\rchi(P)\ge\rchi(Q), \phi(Q)=a\big] \Rightarrow \phi(P)=a.$$
\end{defin}

{\color{black}Next proposition identifies the scfs $\phi: \cP\to \{a, b\}\subseteq A$ that are $(\rchi, a)$-monotone. Such scfs are all and only the canonical scfs corresponding to subsets $C$ of the image set $\rchi(\cP)\big(\subseteq L\big)$ which are SOC with respect to the poset consisting of $\rchi(\cP)$ endowed with the restriction of the ordering of the poset $L$.

\begin{prop}\label{39}
Let $\rchi:\cP\to L$ be a character, $L$ being a poset with respect to a partial order $\ge$.  For a scf   $\phi: \cP\to \{a, b\}\subseteq A$,  $\phi$ is $(\rchi, a)$-monotone  if and only if $\phi=\phi_{(\rchi, C)}$ for some  SOC subset $C$ of the poset $\big(\rchi(\cP), \ge\big)$. When  such  $C$ exists, it is unique.
\end{prop}
\dimo

$\Rightarrow$

Assume  $\phi$ is $(\rchi, a)$-monotone, and consider the set
 $$ C=C_\phi :=\big\{x\in L:   x=\rchi(P) \mbox{ for some }   P\in\cP\mbox{ such that } \phi(P)=a \big\}.$$ 
 This is the desired SOC subset of $\big(\rchi(\cP), \ge\big)$.
 
 Indeed:
 
 -- by $(\rchi, a)$-monotonicity $C$ is SOC in the poset  $\big(\rchi(\cP), \ge\big)$;
 
 -- by $(\rchi, a)$-monotonicity we have $\phi(P)=a\Leftrightarrow \phi_C(P)=a$;
 
 -- $C$ and $C'$ subsets of $\big(\rchi(\cP), \ge\big)$ with $\phi_C=\phi_{C'} \Rightarrow C=C'$.
 
 $\Leftarrow$
 
  Suppose $C$ is a SOC subset of $\big(\rchi(\cP), \ge\big)$, and $\phi=\phi_C$. Assume $\big[P,Q\in \cP,\,\rchi(P)\ge\rchi(Q), \phi(Q)=a\big]$. Since $\phi=\phi_C$, the condition $\phi(Q)=a$ means, by Definition \ref{forma canonica}, that $\rchi(Q)\in C$. By super order closedness of $C$, also $\rchi(P)\in C$, hence $\phi(P)=\phi_C(P)=a$.
\eproof}

\subsection{The character of two-valued non-manipulable social choice functions}

The canonical scf in Definition \ref{forma canonica} becomes the functional form of the two-valued non-manipulable scfs if we suitably define  the character function. To do this we preliminarily have to introduce  a poset.

\begin{defin}\label{312}
We denote by $\ccD$ the set   consisting of triples $(S,T, \pi)$ where $S, T$ are disjoint subsets of $V$, and $\pi$ is a partial profile whose domain is $T$.
We endow 
  the set $\ccD$  with  the  partial order $\le$$_{{\footnotesize\ccD}}$ defined as follows.\footnote{Equivalently one can  define the relation $(S,T, \pi)\le$$_{\footnotesize \ccD}$$(S',T',\pi')$ by requiring the validity of the conditions $S\subseteq S'$, $T\setminus S'\subseteq T'$, and   $v\in T\setminus S'\Rightarrow \pi_v=\pi'_v.$ }
  
  \medskip
\quad\qquad$(S,T, \pi)\le$$_{\footnotesize \ccD}$$(S',T',\pi')$ means that $S\subseteq S'$, $S\cup T\subseteq S'\cup T'$, and   $v\in T\cap T'\Rightarrow \pi_v=\pi'_v.$
\end{defin}

The proof that the relation $\le$$_{{\footnotesize\ccD}}$ is a partial order is  given in the Appendix. We  now  present the character of the two-valued wGSP scfs.

\begin{defin}\label{carattere 2-valued}
The character function for the two-valued weakly group strategy-proof scfs is the function $\rchi:\cP\to \ccD$ defined as $$\rchi (P)= \big(D(a,P), I(P), (P_v)_{v\in I(P)}\big).$$
\end{defin}

A direct verification shows that $P\overset a\ge Q\Leftrightarrow \rchi(P)\ge$$_{{\footnotesize{\ccD}}}$\,$\rchi(Q)$. Hence, we can reformulate Theorem \ref{49} as follows.

\begin{theo}\label{49bis}
 Let $a, b$ be two distinct elements of $A$. Consider a
  scf $\phi:\cP\to \{a, b\}$ defined over a quasi--cartesian restricted domain. Then, $\phi$    is  wGSP   if and only if   $\phi$   is   $(\rchi, a)$-monotone.\end{theo}

\bigskip
We present below the canonical functional form of the non-manipulable two-valued scfs.

\begin{theo}\label{main1}
Let $a, b$ be two distinct elements of $A$.  Assume $\cP$ is a quasi--cartesian restricted domain.
 Let $\rchi$ be the character introduced in Definition \ref{carattere 2-valued}.
 Then,  the map $C \mapsto \phi_{(\rchi, C)}$  is a bijection from the set of all SOC subsets of $\big(\rchi(\cP), \le$$_{\footnotesize \ccD}$$\big)$ to the set of all   wGSP scfs from $\cP$ to $\{a, b\}$.
 \end{theo}
\dimo Apply Proposition \ref{39} and Theorem \ref{49bis}.\eproof

{\color{black}Observe that to determine the collective choice corresponding to a profile $P$ we do not have to know the actual profile $P$, but only need to know the components of $\rchi(P)$. Therefore, these
 constitute the {\it primitive} elements  for the formation of the collective decision.}

An element of $\rchi(\cP)$ is a triple $(S, T, \pi)$ consisting of disjoint subsets $S, T$ of agents and a partial $\{a, b\}$-indifference profile $\pi$. The latter is a partial profile $\pi$ such that for every agent $v$ in its domain $T$,  one has that $a$ and $b$ are indifferent according to $\pi_v$. So, we are dealing with a generalized version of the veto pairs considered in \cite{KPS-GEB}. With $W=S\cup T$ , the elements of $\rchi(\cP)$ can be seen as pairs consisting of a veto pair and a $\pi.$ 

Intuitively: the triple $(S, T, \pi)$ acts  similarly to a veto unit, determining a non-manipulable rule for the collective choice and expressing a veto against $b$. Indeed, to block the alternative $b$ (i.e. $a$ is the collective choice) it suffices  that {\it $(S, T, \pi)$ supports $a$, i.e.  the agents of $S$ vote for $a$, and those $v\in T$ that don't move in favor of a, are  adamant at the  $\{a, b\}$-indifferent preference $\pi_v$. }
{\color{black}\begin{oss}\label{JME}
{\sl In \cite{KPS-JME} Basile et al. also identify a functional form of the wGSP scfs. They are all and only the scfs of $\psi$-type. The definition of $\psi$-type functions relies on an extension of the concept of committee,\footnote{We recall that a committee   is  a nonempty family of coalitions of agents which is closed under supersets. See subsection \ref{strict preferences}, {\it infra}.} and one needs a well ordered collection of extended committees in order to define a $\psi$-type scf $\phi$. The social choice  $\phi(P)$ will  then be the choice of the first extended committee  which is not indifferent, if there is one; otherwise it will be the the alternative  associated to a profile of unanimous indifference between the alternatives $a, b$. As we see this kind of representation is quite different from the canonical functional form. Moreover does not guarantee uniqueness of the representation. Finally, in \cite{KPS-JME} the domain is universal. Theorem \ref{main1} is a notable improvement of the representation theorem proved in \cite{KPS-JME}.
}
\end{oss}
}

\bigskip
When $V$ 
 and A are finite, $\rchi(\cP)$ is finite too, and since SOC subsets of $\rchi(\cP)$ are in a one-to-one correspondence with antichains of $\rchi(\cP)$ (due to Proposition \ref{finito}), the canonical functional form can be expressed in terms of sets of incomparable elements of $\big(\rchi(\cP), \le$$_{\footnotesize \ccD}$$\big)$. 

According to the following representation theorem, rules for a non-manipulable collective choice between two alternatives $a$ and $b$ can be formed uniquely in the following way: with reference to
some incomparable\footnote{Note that incomparability in the poset $\big(\rchi(\cP), \le$$_{\footnotesize \ccD}$$\big)$ may be thought of as  a form of independence in vetoing.} generalized vetoers $(S_1, T_1, \pi^{(1)})$, $(S_2, T_2, \pi^{(2)})$, $(S_3, T_3, \pi^{(3)}),\dots$ the rule selects $a$ if and only if at least one of the $(S_i, T_i, \pi^{(i)})$ supports $a$.

\begin{theo}\label{main1bis}
 Suppose $V$
and A are finite, $\cP=\times_{v\in V} \cW_v$, and $\rchi$ is the character of Definition \ref{carattere 2-valued}. Then,
 there is a bijection  $\bDelta \mapsto \phi_{\bDelta}$  from the set of all antichains $\bDelta$ of $\big(\rchi(\cP), \le$$_{\footnotesize \ccD}$$\big)$ to the set of all  wGSP scfs from $\cP$ to $\{a, b\}$. For  an antichain $\bDelta$ the scf $\phi_{\bDelta}$ is defined as follows
$$
\phi_{\bDelta}(P)= \left\{ 
\begin{array} {ll}
a , & \mbox { if } \exists\,\, (S, T, \pi)\in\bDelta: (D(a,P), I(P), (P_v)_{v\in I(P)})\ge_{\footnotesize \ccD} (S, T, \pi)   \\
b , & \mbox { otherwise }  \\
\end{array}
\right. .$$
\end{theo}

\begin{oss}\label{ilnulla}
\sl
{\color{black}
Hagiwara and Yamamura (2020) prove the following result, which they call closed characterization,  for wGSP scfs with range of cardinality two within a larger set of alternatives. 
\begin{theo}\label{nulla} \cite[Theorem 2, and Corollary 1]{HY}
Let $\phi$ be a scf with range $\{a, b\}$. Then, $\phi$ is wGSP if and only if it is an upper set rule 
$g^{\cal Q}$.
\end{theo}
To define an upper set rule one needs a set ${\cal Q}\subseteq \cP$ of profiles with the property of being an {\it \lq\lq upper set with respect to $\underset\sim\prec_{(b,a)}$\rq\rq}  in the sense defined in \cite[page 662]{HY}.

Then the definition of the rule is the following: 

$$
g^{\cal Q}(P)= \left\{ 
\begin{array} {ll}
a , & \mbox { if } P\in {\cal Q}   \\
b , & \mbox { if } P\notin {\cal Q}   \\
\end{array}
\right. .$$
There are  substantial differences between Theorem \ref{nulla} and our results. Since Hagiwara and Yamamura assume that $V$ is finite, we can  compare Theorem \ref{nulla} and  Theorem \ref{main1bis}. 


 
 We observe that
 
 -- an upper set rule identifies the collective choice corresponding to a profile $P$ by verifying if $P$ belongs to the upper set $\cal Q$. 
 
 --  the knowledge of $\cal Q$ is required to implement this upper set rule.
 

\bigskip
Our characterization does not need a large set of profiles exhausting the set of profiles that will select $a$. First of all,
we just refer to the primitives which are necessary to the formation of the collective choice. These are: the sets $D(x,P)$ of agents in favor of  $x\in\{a,b\}$, the sets $I(P)$ of agents which are indifferent  and, for  each agent $v\in I(P)$, 
the particularly chosen $\{a, b\}$-indifference preference  $P_v$.
Second, a scf is identified by means of a small number of incomparable generalized vetoers,\footnote{The number of 
 incomparable elements within a SOC set $C$ is much smaller than the the number of elements of $ C$ }
   each leading the collectivity to choose $a$. Then  the value of $\phi(P)$ comes from the comparison of the primitives of the profile $P$ 
%
with the vetoers.

To conclude this Remark, we observe that the notion of   upper set with respect to $\underset\sim\prec_{(b,a)}$ can be framed within our more general notion of SOC subset of a poset. Due to this,  Theorem \ref{nulla} can also be seen as a corollary of our Theorem \ref{main1}. Moreover, Hagiwara and Yamamura assume, more restrictively, that $\cP$ is a cartesian restricted domain  with identical factors. 

}
\end{oss}

\bigskip

\section{Anonymous social choice functions}\label{ancase}
In this section we suppose that the set  $V$ of agents is finite, and $\cP=\cW^V$, with $\cW\subseteq\cW(A)$.
Let $\cccfi_{AN}$ consists of all scfs $\phi:\cP\to\{a, b\}\subseteq A$ which are non-manipulable (in this case one can refer to ISP or wGSP equivalently, see Remark \ref{ispwGSP}) and anonymous.

\begin{defin}
A scf $\phi$ is
{\bf anonymous} if,  for every profile $P$ and for every permutation $\sigma$ of $V$, it results 
\,\, $\phi(P)=\phi(P\circ \sigma)$, where $P\circ \sigma$ is the profile in which the preference of each agent $v$ is $P_{\sigma(v)}.$
\end{defin}


We also assume that there are finitely many  preferences in $\cW$ for which the alternatives $a$ and $b$ are indifferent. Hence, setting $\cI=\{W\in\cW: a\sim_W b\}$ we assume this set is finite, say of cardinality $\tau$ ($A$ need not to be finite; $\tau$ may be zero). Let us enumerate the elements of $\cI$ as follows $\cI=\{W_1, W_2,\dots, W_\tau\}$, and introduce the sets $I_i(P)=\{v\in V: P_v=W_i\}$, for $i=1,\dots,\tau$. Of course the sets $I_i(P)$ partition $I(P)$.

\bigskip
Now we describe the character function we associate to the class  $\cccfi_{AN}$. Let us fix  the number of agents to be $n$, {\color{black}and define $L$ to be the set $L=\{0,1,\dots, n\}^{\tau+1}$. Let us use the notation  $t^+$ for the positive part of a number $t$, namely $t^+=\max\{t, 0\}.$ We  make $L$ a poset by introducing the following partial order.

For $\ccck, \cccell\in L$, define:
$$\ccck\le_L \cccell\Leftrightarrow \sum_{i=1}^\tau (k_i-\ell_i)^+\le \l_0-k_0.$$

\medskip
It is straightforward to see that $\ccck\le_L \cccell$ is equivalent to $k_0+\sum_{i\in J} k_i\le \ell_0+\sum_{i\in J} \ell_i$ for every subset $J$ of $\{1,\dots, \tau\}$. 

Indeed, if  $J$ is a subset of $\{1,\dots, \tau\}$ and we assume $\ccck\le_L \cccell$, then we have 

$\sum_{i\in J} (k_i-\ell_i)^+\le 
\sum_{i=1}^\tau (k_i-\ell_i)^+\le \l_0-k_0,$ proving one implication. For the converse observe that we can write
$ \sum_{i=1}^\tau (k_i-\ell_i)^+=\sum_{i\in J} (k_i-\ell_i)^+  + \sum_{i\notin J} (k_i-\ell_i)^+$ if we set $J$ to be the subset of $\{1,\dots, \tau\}$ consisting of all indices  for which $k_i\ge\ell_i$. Consequently,
$ \sum_{i=1}^\tau (k_i-\ell_i)^+=\sum_{i\in J} (k_i-\ell_i)$, and this gives immediately $\ccck\le_L \cccell$, in virtue of the present assumption.\eproof

\medskip
Having introduced the reference poset, we  present the definition of character  for the  anonymous, weakly group strategy-proof scfs.
\begin{defin}\label{carattere anonimo}
The character function for the two-valued, anonymous, weakly group strategy-proof scfs is the function $\rchi:\cP\to \{0,1,\dots, n\}^{\tau+1}=L$,
defined by 
$$\rchi=(\rchi_0, \rchi_1, \dots, \rchi_\tau)\mbox{ with } \rchi_0(P)=|D(a, P)|,\, \rchi_i(P)=|I_i(P)|\mbox{ for } i=1,\dots, \tau.
$$
\end{defin}

}
We now present the main result of this section.
\begin{theo}\label{anonymous0}
Let $\phi:\cP\to\{a, b\}\subseteq A$ be a scf. Then,
$$\phi\in\cccfi_{AN}\Leftrightarrow \phi \mbox{ is } (\rchi, a)\mbox{-monotone. }$$
\end{theo}
\dimo
Let us suppose that $\phi \mbox{ is } (\rchi, a)\mbox{-monotone. }$ Because of Proposition \ref{39}, we know that $\phi=\phi_C$ for some SOC subset $C$ of the poset $\big(\rchi(\cP), \le_L\big)$.
{\color{black}Since for every profile $P$ and every permutation $\sigma$ of agents the cardinalities 
$|D(a, P\circ\sigma)|,\, |I_i(P\circ\sigma)|$ are respectively the same as
$|D(a, P)|,\, |I_i(P)|$, the character values $\rchi(P)$ and $\rchi(P\circ\sigma)$ are identical. Hence, by definition of
$\phi_C$, we obtain  that $\phi_C$ 
 is  anonymous.} It remains to prove that it is also weakly group strategy-proof. By appealing to Theorem \ref{49}, we can prove that it is almost monotone:

$$P\overset a\ge Q, \phi_C(Q)=a \Rightarrow \phi_C(P)=a,$$ or, equivalently, that

$$P\overset a\ge Q, \rchi(Q)\in C \Rightarrow \rchi(P)\in C.$$ 

The assertion will follow from showing that 

\medskip
CLAIM 1: 
$P\overset a\ge Q\Rightarrow \rchi(P)\ge_L\rchi(Q)$, 

and from the super order closedness of $C$. CLAIM 1 is proved in  Appendix \ref{anonymous1}.

For the converse, let us assume that $\phi\in\cccfi_{AN}$.  In  Appendix \ref{anonymous2} the following claim is proved.

\medskip
CLAIM 2:  The set $C=\{\rchi(P): \phi(P)=a\}$ is SOC in $(\rchi(\cP), \le_L)$.

\medskip
As a consequence, we have that $\phi=\phi_C$, i.e. $\phi(P)=a \Leftrightarrow \phi_C(P)=a$. Indeed the implication $\phi(P)=a \Rightarrow \phi_C(P)=a$ is true by definition of $C$, whereas the converse $\phi(P)=a \Leftarrow \phi_C(P)=a$  derives from the anonymity of $\phi$ together with the definition of the character function.
Finally, Proposition \ref{39} applies.
\eproof

Because of Proposition \ref{finito}, and Proposition \ref{39} we have the following corollary of the above Theorem \ref{anonymous0}.
\begin{corol}\label{anonymous3}
 Suppose the set $\cP$ of the feasible profiles is $\cW^V$. Let  $a, b$ be two alternatives belonging to the set $A$. Assume that  $V$,  as well as the preferences in $\cW$ for which the alternatives $a$ and $b$ are indifferent,
 are finite. Then,
 there is a bijection  $\bDelta \mapsto \phi_{\bDelta}$  from the set of all antichains $\bDelta$ of $\big(\rchi(\cP), \le$$_{\footnotesize L}$$\big)$ to the set $\cccfi_{AN}$ of all  anonymous, weakly group strategy-proof  scfs with values in $\{a, b\}$. For  an antichain $\bDelta$ the scf $\phi_{\bDelta}$ is defined as follows
$$
\phi_{\bDelta}(P)= \left\{ 
\begin{array} {ll}
a , & \mbox { if } \exists\,\, (x_0,x_1,\dots ,x_\tau)\in\bDelta: (|D(a,P)|, |I_1(P)|, \dots,|I_\tau(P)|)\ge_{\footnotesize L} (x_0,x_1,\dots ,x_\tau)   \\
b , & \mbox { otherwise }  \\
\end{array}
\right. .$$
\end{corol}

Hence in the anonymous case for the formation of the collective decision the primitives are the components of a numerical vector. Precisely, the number of agents that vote for $a$, and, for $i=1,\dots,\tau$, the number of agents that express the $\{a, b\}$-indifference preference $W_i$

\begin{ex} {\sl We   illustrate  with an example the role of the character function.

 
  To fix ideas let us suppose that in $\cW$ there are three preferences $W_1, W_2, W_3$ under which $a$ and $b$ are indifferent. The agents are eleven:  $V=\{v_1, \dots, v_{11}\}$. 

In this case it turns out that the appropriate poset $L$ to use consists of the $4$-tuples of elements of  $\{0,1,\dots, 11\}$,
where for $\cccx=(x_0, x_1, x_2, x_3),\cccy=(y_0, y_1, y_2, y_3)\in L$ we define $$\cccx\le \cccy\Leftrightarrow x_0+\sum_{i=1}^3\, (x_i-y_i)^+\leq y_0.$$
The character function is $P\in \cP\mapsto \rchi(P)\in L$ by letting $(\rchi(P))_0$ be the number of agents $v\in V$ that  prefer $a$ to $b$ under $P_v$, and $(\rchi(P))_i$ be the number of agents for which $P_v=W_i$ ($i=1, 2, 3$).

Hence, 
$$\rchi(\cP)=\Big\{\cccx\in L : \sum_{i=0}^3\, x_i\leq 11\Big\}.$$ The elements of $\cccfi_{AN}$
are all and only the $\{a, b\}$-valued scfs $\phi$ defined on $\cP$ which are monotone in the following sense

$$P, Q\in \cP, \phi(Q)=a, \rchi(P)\ge\rchi(Q)\Rightarrow \phi\,(P)=a,
$$
or, equivalently, all and only the scfs $\phi_{(\rchi, {\footnotesize C})}$ where $C$ runs over the SOC subsets of $\big(\rchi(\cP),\ge\big)$.

Since the latter is  finite, we can equivalently parametrize $\cccfi_{AN}$ with the minimal elements of $C$ rather than with $C$ itself.  This identification tells us that the character function permits the description of all scfs in simple understandable terms.
For example, if the minimal elements of $C$ are the following three incomparable quadruples 
$\cccx = (2,4,0,3),\, \cccy=(4,1,3,1),\, \cccz=(4,2,4,0)$ of $\rchi(\cP)$, the correspondent anonymous non-manipulable two-valued scf $\phi$ is 

$$\small
\phi(P)= \left\{ 
\begin{array} {ll}
a , & \mbox {if  at least one of the following } \rchi(P)\ge \cccx, \rchi(P)\ge \cccy, \rchi(P)\ge \cccz, \mbox{ holds true,}   \\
b , & \mbox {otherwise. }  \\
\end{array}
\right.$$

In words: the corresponding scf is the one that selects $a$  
in each  profile\footnote{Like the profile $P$ whose primitives are $(3,3,1,4)$.} that shows at least the same support for the alternative $a$ than either $\cccx$, or $\cccy$, or $\cccz$. Otherwise\footnote{Like for the profile $Q$ whose primitives are $(4,1,1,2)$.}  the collective choice is $b$. The profile $\cccx= (2,4,0,3)$, for example, simply means that two votes are for $a$, four agents choose $W_1$, three choose $W_3$, and no one chooses $W_2$. Whereas to support alternative $a$ no less than $\cccx$ means that  a profile $P$ forms under which:
\begin{itemize}
\item[--] at least two votes are for $a$,
 
\item[--]  the number of agents that are either in favor of $a$ or express the preference $W_j$ is at least $2+x_j$,
  
\item[--]   the number of agents that are either in favor of $a$ or express one of the two preferences $W_i, W_j (i\neq j),$ is at least $2+x_i+x_j$,
   
\item[--]   the number of agents that are either in favor of $a$ or choose one of the three preferences $W_i$ is at least nine.\eproof\end{itemize}
}\end{ex}

 \section{The unifying role of character and canonical functional form}\label{unifying}
 {\color{black}

\bigskip
 Let us denote by $\cccfi$ the class of all wGSP scfs $\phi: \cP\to \{a, b\}$, under the basic assumptions that we have adopted, i.e. 
 
 -- $\cP$ is a quasi-cartesian restricted domain,
 
 -- the distinct alternatives $a, b$ belong to a larger set $A$ of alternatives.

 In Section \ref{tre},  for the class $\cccfi$,  we have defined a character  with the help of which in Theorem \ref{main1}   we have provided the canonical functional form for all the scfs in $\cccfi$.
 
 In the same way, in Section 4, we have defined a character for the class $\cccfi_{AN}$  and provided the canonical functional form for all the scfs in this class.

 In each of the respective cases above, if $\rchi$ is the character corresponding to the respective class, the scfs in that class are  all and only the functions $$(\star)\qquad \{\phi_{(\rchi, C)}: C \mbox { is a SOC subset of } \rchi(\cP)\}.$$
 
  In other words these two classes admit the same functional representation  introduced in Definition \ref{forma canonica}. The character acts as a parameter that distinguishes the classes, and the SOC set $C$ acts as a parameter that distinguishes the scfs within a class.

 The purpose of this section is to consider other classes of scfs, provide the corresponding character functions and  describe 
 the corresponding class in the fashion described by $(\star)$ above. 

We will first consider the classes: 
  
 \begin{itemize}
  \item [$\cccfi^\cS$,] the class of all wGSP scfs $\phi: \cP\to \{a, b\}$, under the  assumption that  
$\cP=\cS(A)^V$ is the strict universal domain;
 \item [$\cccfi^{bi}$,] 
 the class of all wGSP scfs $\phi: \cP\to \{a, b\}$, under the  assumptions that $A= \{a, b\}$;
 \item [$\cccfi_{AN}^{bi}$,] the subclass of $\cccfi_{AN}$ under the  assumptions that $A= \{a, b\}$.
  \end{itemize}
 
 As a further exemplification of  the unifying possibilities  offered by  the structure of the function 
 $\phi_{(\rchi, C)}$, we also consider   the subclass
 $\cccfi_{\rm s}$ of $\cccfi$ consisting of all strongly group strategy-proof scfs (see  Definition 4 in  Barber\`a et al. \cite{BBM}). 
 
The possibility of unifying by means of $(\star)$ the description of so many relevant classes, is the reason why we have proposed the name canonical for such scfs.

\subsection{Strict preferences.}\label{strict preferences}
 We recall that a committee (also named {\it simple game}  in the Game Theory literature)  is  (see  \cite{Peleg}), a nonempty family of coalitions of agents which is closed under supersets. We also recall that a \lq\lq voting by committee\rq\rq\, is a scf $P\mapsto \phi_C(P),$  where $C$ is a committee of agents, and    
$\phi_C(P)=a\overset{{\rm def}}\Longleftrightarrow D(a, P)\in C.$

When $V$ is finite, and $A=\{a, b\}$ Larsson and Svensson  proved the following \cite[Theorem 2]{LS}: {\it  \lq\lq a voting rule defined for all strict
preference profiles, is strategy-proof \footnote{They mean individually, but this is  the same as wGSP, as we saw previously.} and onto if and only if it is voting by committees.\rq\rq}

Let us observe that this functional form characterization  is canonical. Indeed, it is such by using the character function $\rchi$ introduced in Definition \ref{carattere 2-valued} with reference to the strict universal domain. Let us suppose that  $\cP$ is the strict universal domain.
The second and the third components of the character  are, respectively, the empty set and the empty profile. Consequently we can identify 
 $\big(\rchi(\cP), \le$$_{\footnotesize \ccD}$$\big)$ with the power set of $V$ ordered by the set inclusion and the concept of super order closed subset of $\rchi(\cP)$ gives back either {\it stricto sensu} a committee
  or the empty set or the entire power set. Hence Theorem \ref{main1} gives

\begin{corol} 
Let $V, A$ be arbitrary and $a, b$ two distinct elements of $A$. Suppose that $\cP$ is the strict universal domain. Let $\rchi:\cP\to 2^V$ defined by $\rchi(P)=D(a,P)$.

 Then, $\cccfi^\cS$ consists of  all and only the functions $P\mapsto \phi_{(\rchi, C)}(P),$ \, where $C$ is a subset of  agents closed under supersets.
\end{corol}
The above corollary generalizes \cite[Theorem 2]{LS}.
}

\subsection{Two alternatives only}\label{sezione5}
In this subsection we assume $A=\{a, b\}$, i.e.  agents express preferences only about the realizable alternatives. This is the framework of Basile et al. \cite{KPS-GEB} in which  finiteness of $V$ and universality of the domain $\cP$ is assumed. We do not assume these hypothesis in the following. 

In this  framework Theorems \ref{main1} and \ref{main1bis}  can be simplified on the basis of the following observations.
Since, given a profile $P$, the partial profile $(P_v)_{v\in I(P)}$ is uniquely determined, we  identify the triple $\big( D(a, P), I(P), (P_v)_{v\in I(P)}\big)$  with the pair 
{\color{black}$\big(D(a, P), \overset\sim D(a, P)\big)$.  The role of the poset $L$, instead of $\ccD$, can be played by the simpler poset 
$\cV$ consisting of the set $\{(S,W)\in 2^V\times 2^V: S\subseteq W\}$ endowed with the componentwise set inclusion as partial order $\le$$_{\footnotesize \cV}$. \footnote{The poset $\cV$ is clearly embeddable into $\ccD$.} 
Likewise the role of the character of Definition \ref{carattere 2-valued} can be played by a simpler one. We mean the character  $\rchi_{bi}:\cP\to\cV$ defined as $\rchi_{bi}(P)=\big(D(a, P), \overset\sim D(a, P)\big)$.

We recall that $\cV$ has been introduced in Basile et al. in \cite{KPS-GEB}. Its elements were named {\it veto pairs}.
}

Now, Theorems \ref{main1} and \ref{main1bis} specialize as follows in case $A=\{a, b\}$.
\begin{theo}\label{corollario 52} {\rm (Theorem \ref{main1} in case of two alternatives only). }
Assume $A=\{a, b\}$. On the quasi--cartesian restricted domain $\cP$, the class $\cccfi^{bi}$ consists of
all and only the functions $\phi_{(\rchi_{bi},C)}$, with   $C$  SOC set of veto pairs.
\end{theo}

The above theorem is new with respect to \cite{KPS-GEB}, since  $V$ is arbitrary and $\cP$ need not to be the universal domain.

The Corollary below is  the {\bf veto pairs representation} of non-manipulable scfs provided in \cite[Theorem 3.3]{KPS-GEB}, limitedly to the universal domain. Again we have a functional form characterization of canonical type.

{\color{black}
\begin{corol}\label{corollario 53} {\rm (Theorem \ref{main1bis} in  case of two alternatives only). }
 Suppose $V$ 
 is finite, $A=\{a, b\}$, and $\cP=\times_{v\in V} \cW_v$. Then,
  there is a bijection  $\cSigma \mapsto \phi_{\cSigma}$  from the set of all antichains $\cSigma$ of $\big(\rchi_{bi}(\cP), \le_{\footnotesize{\cV}}\big)$ to $\cccfi^{bi}$.
  For  an antichain $\cSigma$ the scf $\phi_{\cSigma}$ is defined as follows
$$
\phi_{\cSigma}(P)= \left\{ 
\begin{array} {ll}
a , & \mbox { if } \exists\,\, (S, T)\in\cSigma: (D(a,P), \overset\sim D(a,P))\ge_{\footnotesize \cV} (S, T)   \\
b , & \mbox { otherwise }  \\
\end{array}
\right. .$$
\end{corol}

 Example \ref{Esempio 11} marks once more the distinction between assuming or not that $A=\{a, b\}$. Indeed if we consider, with $A\supset \{a, b\}$, the $\{a, b\}$-valued scfs $\phi_{\cSigma}$ defined above are wGSP but do not exhaust all $\{a, b\}$-valued wGSP scfs. The function of Example \ref{Esempio 11} cannot be represented as a 
$\phi_{\cSigma}$. Its functional form  comes from Theorem \ref{main1bis} and cannot come from Corollary \ref{corollario 53}.
This is so because  $\phi_{\cSigma}$, differently from $\phi_{\bDelta}$ in Theorem \ref{main1bis}, takes only care of which agents $v$ are indifferent and not even of their  preference $P_v$.

 This tells us that investigating the setting with $A\supseteq \{a, b\}$ is not obvious from \cite{KPS-GEB}.

\def\bG{\mbox{\boldmath $G$}}

\bigskip
Similar considerations to the previous ones can be carried out in the anonymous case with  exactly two alternatives. The poset $L$ of Section \ref{ancase} becomes $\{0,1, \dots, n\}\times \{0,1, \dots, n\}$,   endowed with the order relation:
$\cccx\le_L\cccy\Leftrightarrow x_0\le y_0, $ and $x_0+x_1\le y_0+y_1$. 

Assume $\cP$ is the universal domain, so that we are in the framework of \cite{KPS-GEB}.
Let 
$\mathbb{N}_0$ be the set $\{0, 1, 2,\dots \}$.

It is straightforward to verify that the set $\rchi(P)=\{(x_0,x_1)\in \mathbb{N}_0\times\mathbb{N}_0 : x_0+x_1\le n\}$ with the order $\ge_L$ is  order isomorphic to the poset (introduced in \cite[Subsection 4.1]{KPS-GEB}) $\big(G,\ge_G\big)$ where
$$G=\{(x_0,x_1)\in \{0,1,\dots,n\}^2: x_0\le x_1\}, \mbox{ and the order is } \cccy\ge_G\cccx\Leftrightarrow y_i\ge x_i, \, i=0,1.$$ Hence, Corollary \ref{anonymous3} says that the veto cardinality representation \cite[Theorem 4.3]{KPS-GEB} by of Basile et al. is a functional form characterization of $\cccfi^{bi}_{AN}$ of canonical type.

We conclude this section by remarking that  even the quota majority method (see \cite[Corollary p. 63]{M}) is a functional form characterization of canonical type. Indeed, in the case  of strict preferences only,
the poset $L$ of Section \ref{ancase} becomes $\{0,1, \dots,  n\}$, endowed with the usual order. The character is $\rchi(P)= |D(a,P)|$. The fact that all and only the non-manipulable scfs are the functions $\phi_C$ with $C$ SOC subset of $\{0,1, \dots, n\}$, gives back the quota majority method.
}

 

\subsection{Strong group strategy-proofness} \label{sezione6}

In this section we exemplify how the character function approach applies to the result by Barber\`a et al. \cite{BBM} concerning the functional form of strongly group strategy-proof scfs.
Notice that their results require the assumption about $\cP$ they refer to as {\it \lq\lq minimal assumption\rq\rq}:

$\cP=\times_{v\in V}\cW_v$ with $\cW_v\subseteq \cW(A)$, and every $\cW_v$ contains at least three preferences $W^i\in \cW(A)$ for $i=\sim, a, b$ (depending on $v$) such that in $W^\sim$  $a,b$ are indifferent, in $W^a$  $a$ is preferred to $b$, in $W^b$ $b$ is preferred to $a$.


Let us recall that a scf is strongly group strategy-proof when it is immune to weak manipulation. In contrast to Definition \ref{manipulation} of strong manipulation the weak manipulation is defined as follows.
{\color{black}
\begin{defin}\label{weak manipulation}
Let $\phi$ be scf.
 We say that a coalition $D$ can weakly manipulate a profile $P\in \cP$   under $\phi$  if there is another profile $Q\in\cP$ such that 

\begin{itemize}
\item every agent $v$ in $D^c$ has the same preference in both $P$ and $Q$, \, i.e. $P_v=Q_v$; 
\item  for every agent $v$ in $D$,     $\phi(Q)\meglio_{P_v} \phi(P)$, and at least for one agent $v_o$ of $D$,  $\phi(Q)\succ_{P_{v_o}} \, \phi(P)$.
\end{itemize}
\end{defin}
}
i.e. all members of the manipulating coalition are not worse off and at least one is better off.

 Let $\widehat\cU$ be  the set  $\{P\in\cP: I(P)=V\}$, i.e. the set of all feasible profiles under which all agents, unanimously, are indifferent between alternative $a$ and alternative $b$. 
 
 To fix  ideas we consider the case that there are at least three agents in $V$ \footnote{There is a difference in case $|V|=2$, but for our illustrative purposes the  case $|V|>2$ is sufficient.}.

It is known from \cite[Theorem 3]{BBM} that a scf $\phi$ with range of cardinality two is strongly group strategy-proof  if and only if it is, with a suitable choice of a subset $\cU$ of $\widehat\cU$, either a 

{\bf veto for b}:
$$ \phi(P)=a \overset{def}\Longleftrightarrow \mbox{ either } D(a, P)\neq\O \mbox{ or } P\in \cU,
$$ 
or a {\bf veto for a}:
$$ \phi(P)=a \overset{def}\Longleftrightarrow \mbox{ either } \big[D(a, P)\neq\O\mbox{ and } D(b, P)=\O\big] \mbox{ or } P\in \cU.
$$
Let $(L,\ge)$ be the poset  (see the figure below) with $L=\{-1,0, 1\}\cup \widehat \cU$ and where $1$ and $-1$ are respectively the maximum and the minimum, and the remaining elements being pairwise incomparable. 

\bigskip
\begin{tikzpicture}
\draw[fill] (0,0) circle [radius=0.1];\node [below left] at (0,0) {$-1$};
\draw[fill] (0,2) circle [radius=0.1];\node [below left] at (0,2) {$0$};
\draw[fill] (0,4) circle [radius=0.1];\node [below left] at (0,4) {$1$};
\draw[fill] (3,2) circle [radius=0.1];
\draw[fill] (4,2) circle [radius=0.1];
\draw[fill] (5,2) circle [radius=0.1];
\draw[fill] (10,2) circle [radius=0.1];
 \draw (6.5,2) ellipse (4cm and .7cm);
\draw  [dashed]  (0,4)  -- (10,2);\draw  [dashed]  (0,4)  -- (5,2);\draw  [dashed]  (0,4)  -- (4,2);\draw  [dashed]  (0,4)  -- (3,2);
\draw  [dashed] (0,4)  -- (0, 0);
\draw  [dashed]  (0,0)  -- (10,2);\draw  [dashed]  (0,0)  -- (5,2);\draw  [dashed]  (0,0)  -- (4,2);\draw  [dashed]  (0,0)  -- (3,2);
\node [left] at (7,2) {$\dots$}; \node [left] at (8,2) {$\dots$}; \node [left] at (9,2) {$\dots$};
\node at (9,3) [above] {$\widehat\cU$};
\node at (5,-1) [below] {The poset $L$};
\end{tikzpicture}

Let us define the character function $\rchi: \cP\to L$ as follows:
 $$
 \rchi (P)= 
 \left\{ 
\begin{array} {ll}
 1,   & \mbox{ if }   D(a,P)\neq\O \mbox{ and } D(b,P)=\O,\\
 0,   & \mbox{ if }   D(a,P)\neq\O \mbox{ and } D(b,P)\neq	\O,  \\
 -1,   & \mbox{ if }   D(a,P)=\O \mbox{ and } D(b,P)\neq\O, \\
 P,   & \mbox{ if }   D(a,P)=\O \mbox{ and } D(b,P)=\O. \\
\end{array}
\right.
 $$
 The SOC subsets $C$ of $(L,\ge)$ are: the empty set;
$L$; the sets $\{0,1 \}\cup \cU$ with $\O\subseteq\cU\subseteq\widehat\cU$; 
the sets $\{1 \}\cup \cU$ with $\O\subseteq\cU\subseteq\widehat\cU$. 
When $C$ is empty $\phi_{(\rchi, C)}$ is constantly $b$.  
When $C=L$ then $\phi_{(\rchi, C)}$ is constantly $a$. For $C=\{0,1 \}\cup \cU$, the scf $\phi_{(\rchi, C)}$ is the veto for $b$ and in the latter case we get the veto for $a$, in this way exhausting the class $\cccfi_{\rm s}$ of all strongly group strategy-proof scfs with range of cardinality at most two. In other words the canonical functional characterization applies to $\cccfi_{\rm s}$ too.

\section{Conclusions}

{\color{black}In this study we have considered a social choice model with an arbitrary set of agents $V$,  and a set $A$ of at least two alternatives. The collective choice is limited to either $a$ or $b$ ($a,b\in A$),  and is based upon a preference profile of the agents who also consider the other alternatives in $A$. Indifference is permitted. The non-manipulability notion adopted is the weak group strategy-proofness.

We have provided the functional form for each of the following classes of $\{a, b\}$-valued scfs $\phi$:

- all non-manipulable scfs, 

- all  non-manipulable  and anonymous scfs (here $V$ is a finite set of agents),
 
 - all non-manipulable  scfs defined on the strict universal domain,

-  all non-manipulable  scfs, under the  assumptions that $A= \{a, b\}$,

-  all non-manipulable anonymous, under the  assumptions that $A= \{a, b\}$,

- all strongly group strategy-proof scfs.

\bigskip
In each of these cases, for an appropriate poset $(L, \ge_L)$, an appropriate function $\rchi$ from the set $\cP$ of feasible profiles to $L$, and appropriate subset $C$ of $L$, we have shown that
$$\phi(P)=a \mbox{ if and only if } \rchi(P)\in C,$$
The set $C$ is required to have the following properties 

(1) $C$ is contained into the $\rchi$-image of $\cP$,

(2) $\rchi(Q)\in C$, and $\rchi(P)\ge_L \rchi(Q) \Rightarrow \rchi(P)\in C$.

\bigskip
The distinction of the different cases is marked by a specific pair $L, \rchi$ for each case. So, the different character of the above types of social choice functions is captured by $\rchi: \cP\to L$, and for this reason we named it character function. 

 The character function  with the help of the set $C$, describes how a given scf selects the alternative,  i.e. gives functional forms characterizations of a particular class. Such characterizations share  the same mathematical structure. For this unifying effect  we have spoken of canonical functional forms.  
 
 When the number of agents and the number of alternatives are finite, in the parametrization of $\phi$ we can replace $C$ with the set of its $\ge_L$-minimal elements. This forms a  set (of incomparable elements in $L$) whose cardinality is much smaller than the cardinality of $C$. This way of functional form expression of a social choice function is not only a characterization per se, but also an implementable algorithm. Indeed, the identification of $\phi(P)$  is based on the comparison of  $\rchi(P)$ with the minimal elements parametrizing $\phi$. It is not necessary to know entirely the profile $P$. For the formation of the collective choice, $\rchi(P)$ gives the primitives one needs.
 }

\bigskip

 \section{Appendix}
 
In this section we present the proof of  Proposition \ref{finito}, verify that the relation  $\le$$_{{\footnotesize\ccD}}$ of  Definition \ref {312} is a partial order, and prove the two claims contained in  the proof Theorem \ref{anonymous0}. 
 
 \subsection{The proof of Proposition \ref{finito}}
 
 A lemma precedes the proof of Proposition \ref{finito}. Assume here that $(L,\ge)$ is a given poset.
 
  \begin{lem}
 Let $F$ be a nonempty finite subset of $L$. Then, every element of $F$ is greater than a minimal element of $F$. In other words: $m(F)\neq \O$.
 \end{lem}
 \dimo
 We show that for every $f_0\in F$ there is $x\in m(F)$ such that $f_0\ge x$. If $f_0$ is not minimal itself, the set $F_0=\{x\in F\setminus \{f_0\}: f_0\ge x\}\neq\O.$ Given an element $f_1\in F_0$, if it  belongs to $m(F)$, we are done. Otherwise the set $F_1=\{x\in F\setminus \{f_0, f_1\}: f_1\ge x\}\neq\O.$
 
 Given an element $f_2\in F_1$, if it  belongs to $m(F)$, we are done. Otherwise the set $F_2=\{x\in F\setminus \{f_0, f_1, f_2\}: f_2\ge x\}\neq\O\dots$ Since $F$ is finite, this procedure stops, finding a minimal element of $F$ smaller than the starting point $f_0$.
 \eproof

{\bf Proposition \ref {finito}}  The antichains of a finite poset $L$ are in a one--to--one correspondence with the SOC subsets of $L$.

\dimo
We show that the map $C\mapsto m(C)$ which associates to every SOC subset $C$ of $L$ the antichain consisting of its minimal elements is both
\begin{itemize}
\item injective: $\big[m(C_1)=m(C_2),$ where $C_1, C_2$ are SOC subsets of $L\big]\Rightarrow C_1=C_2,$
\item[ ]and
\item surjective: every antichain $X$ of $L$ is the image $m(C)$ of some SOC subset $C$ of $L$.
\end{itemize}

Assume $m(C_1)=m(C_2),$ where $C_1, C_2$ are SOC subsets of $L$, by symmetry it is enough proving that $C_1\subseteq C_2$ to achieve injectivity. So, take $x\in C_1$. Since $L$ is finite, we can appeal to the previous lemma to find a minimal element $m$ of $C_1$ such that $x\ge m$. Since $m(C_1)=m(C_2),$ the element $m$ belongs to $C_2$ also. Because $C_2$ is SOC, then it contains $x$ too, as desired.

To achieve surjectivity, take a antichain $X$ and set $C$ to be the SOC subset of $L$ defined as follows 
$$ C=\{x\in L: \exists m\in X \mbox{ such that } x\ge m\}.$$ We show that $X\subseteq m(C)\subseteq X.$

For the first inclusion, take $x\in X$. The element $x$ also belongs to $C$, more precisely we have to show that it is also a minimal element of $C$, i.e. from $\big[ x\ge y,\,y\in C\big]$ we must be able to derive that $y=x$. By definition of the set $C$, we get the existence of $m\in X$ such that $y\ge m$. So by transitivity, for the two elements $x, m\in X$ we have $x\ge m$. Since $X$ is an antichain, we get $x=m$. Summarizing $m=x\ge y\ge x=m$ and by antisymmetry we get $y=x$, as desired.

For the second inclusion, let us take $x\in m(C)$. Being $x$ an element of $C$ by definition, we can take $m\in X$ with $x\ge m$. Being $C$ a superset of $X$, we have $\big[ x\ge m, \,$ and $ x, m\in C\big]$. Being $x$ a minimal element of $C$, the elements $x$ and $m$ are identical so $x\in X$.
\eproof
 
 \subsection{The relation $\le$$_{{\footnotesize\ccD}}$ introduced in Definition \ref{312} is a partial order }
 
Transitivity.  Let us suppose that
$(S,T, \pi)\le$$_{\footnotesize \ccD}$$(S',T',\pi')$, \,
and \,
 $(S',T', \pi')\le$$_{\footnotesize \ccD}$$(S'',T'',\pi'')$.  

This two conditions  mean respectively

-- $S\subseteq S'$, $S\cup T\subseteq S'\cup T'$, and   $v\in T\cap T'\Rightarrow \pi_v=\pi'_v,$

-- $S'\subseteq S''$, $S'\cup T'\subseteq S''\cup T''$, and   $v\in T'\cap T''\Rightarrow \pi'_v=\pi''_v,$

\bigskip
Hence the two conditions $S\subseteq S''$, $S\cup T\subseteq S''\cup T''$ that are part of the relation $(S,T, \pi)\le$$_{\footnotesize \ccD}$$(S'',T'',\pi'')$, \,
 that we want to prove are immediate. So we only have to verify that $v\in T\cap T''\Rightarrow \pi_v=\pi''_v$.
 Observe that if $v\in T\cap T''$, then either $v\in  S' $ or $v\in T'$. But the first possibility would give $v\in S''$ which is impossible since $S''$ and $T''$ are disjoint. So the only possibility is $v\in T'$. Consequently we have
 $v\in T\cap T'$, and  $v\in T'\cap T''$. Hence, $\pi_v=\pi'_v=\pi''_v$. 
\bigskip
 
 Antisymmetry. Let us suppose:
$(S,T, \pi)\le$$_{\footnotesize \ccD}$$(S',T',\pi')$, \,
and \,
 $(S',T', \pi')\le$$_{\footnotesize \ccD}$$(S,T,\pi)$.  Namely that
 
-- $S\subseteq S'$, $S\cup T\subseteq S'\cup T'$, and   $v\in T\cap T'\Rightarrow \pi_v=\pi'_v,$

-- $S'\subseteq S$, $S'\cup T'\subseteq S\cup T$, and   $v\in T'\cap T\Rightarrow \pi'_v=\pi_v.$

So $S=S'$ and $S\cup T= S'\cup T'$ are immediate.
Since $S$ and $T$ are disjoint (and $S'$ and $T'$ as well),  one gets that  $T$ and $T'$ are identical and the partial profiles $\pi$ and $\pi'$ too.

 \subsection{CLAIM 1 of Theorem \ref{anonymous0}     }\label{anonymous1}
We have to prove the following  implication: $P\overset a\ge Q\Rightarrow \rchi(P)\ge_L\rchi(Q)$.

\dimo
Let us assume that $P\overset a\ge Q$. 
This is equivalent to:
$$(1)\quad D(a, Q)\subseteq D(a, P), \,\, D(b, P)\subseteq D(b, Q), \,\,\mbox{ and } v\in I(P)\cap I(Q)\Rightarrow P_v=Q_v.$$

From the two partitions $\{D(a, Q), I(Q), D(b, Q)\}$, and  $\{D(a, P), I(P), D(b, P)\}$, of $V$, due to $(1)$, 
we have the partitions


$$ (2) \quad V= D(a, Q) \overset \cdot\cup [D(a, P)\setminus D(a, Q)]\overset \cdot\cup [I(P)\cap I(Q)]\overset \cdot\cup[D(b,Q)\cap I(P)]\overset \cdot\cup D(b,P),
$$ and
$$ (3) \quad V= D(a, Q) \overset \cdot\cup[I(Q)\cap D(a,P)]\overset \cdot\cup[I(Q)\cap I(P)]\overset \cdot\cup[D(b, Q)\setminus D(b,P)]\overset \cdot\cup D(b,P).
$$ 

Let us set: $D=I(Q)\cap I(P),\, E=[D(a,P)\setminus D(a, Q)],\, F=[D(b,Q)\setminus D(b,P)]$. Then, for every $i\in\{1,\dots,\tau\}$, using $(1)$ again, we have:
$$I_i(Q)=[I_i(Q)\cap E]\overset \cdot\cup[I_i(Q)\cap D], \mbox{ from } (2),$$ and
$$I_i(P)=[I_i(P)\cap F]\overset \cdot\cup[I_i(P)\cap D], \mbox{ from } (3).$$
Also notice that $ I_i(Q)\cap D= I_i(P)\cap D$.\footnote{We show the inclusion $\subseteq$. From $v\in I_i(Q)\cap D=I_i(Q)\cap I(P)$, due to $(1)$, we have $P_v=Q_v=W_i$. Hence $v\in I_i(P).$}

\medskip
{\color{black}Now,  because of the latter equation and since the set $D$ is disjoint from both $E$ and $F$, by setting $J=\{i: |I_i(Q)|>|I_i(P)|\}$, we have}

$$\sum_{i=1}^\tau \big[|I_i(Q)|-|I_i(P)|\big]^+=\sum_{i\in J}|I_i(Q)|-|I_i(P)|=\sum_{i\in J}|I_i(Q)\cap E|-|I_i(P)\cap F|\le
$$
$$\le \sum_{i\in J} |I_i(Q)\cap E|\le\sum_{i=1}^\tau |I_i(Q)\cap E|= |I(Q)\cap E|\le |E|=|D(a,P)|-|D(a,Q)|
$$ (again because of $(1)$) which is the same as the desired relation $\rchi(P)\ge_L\rchi(Q)$.
\eproof

\subsection{CLAIM 2  of Theorem \ref{anonymous0}     }\label{anonymous2}
 The claim is: the set $C=\{\rchi(P): \phi(P)=a\}$ is SOC in $(\rchi(\cP), \le_L)$.

\dimo
Let us consider two profiles $P', Q'$ with $\phi(P')=a$, and $\rchi(P')\le_L\rchi(Q')$. We show that 
$\phi(Q')=a$.

For brevity, we set $\rchi(P')=(k_0, k_1,\dots, k_\tau)$, and  $\rchi(Q')=(\ell_0, \ell_1,\dots, \ell_\tau)$. Moreover, we partition the set $\{1,2, \dots, \tau\}$ as $J\cup J^c$ where 
$j\in J\Leftrightarrow k_j>\ell_j.$ Without loss of generality we assume that $J\neq\O$ (the case $J=\O$ is similar and in a sense easier). 
Let us enumerate the elements of $J$ and $J^c$ as follows:
$ J=\{j_1, j_2, \dots, j_\beta\}\qquad J^c=\{i_1, i_2, \dots, i_\alpha\}.
$

We shall introduce two new profiles $P,$ and $Q$, respectively permutations of $P'$, and $Q'$, and show that from $\phi(P)=a$ (due to anonymity) we get $\phi(Q)=a$. Again anonymity will produce the asserted $\phi(Q')=a$.

For convenience we denote the agents as $v_1, v_2, \dots, v_n$, and use the intuitive interval notations:  $\quad V=[v_1, v_n], \quad \{v_{\gamma +1}, v_{\gamma +2}, \dots, v_\delta\}=]v_\gamma, v_\delta]$,\quad and so forth.

Let us consider the sequence of integers $$0\le k_0\le \gamma_1\le\dots\le\gamma_\alpha\le\delta_1\le\dots\le\delta_\beta\le n,$$where: 

$\gamma_r=k_0+k_{i_1}+\dots + k_{i_r}$ for $r=1,\dots,\alpha$, and 

$\delta_r = \gamma_\alpha+ \ell_{j_1}+\dots + \ell_{j_r}$ for $r=1,\dots,\beta$.

Obviously, $$\gamma_\alpha=k_0+\sum_{i\in J^c} k_i, \quad \delta_\beta= k_0+\sum_{i\in J^c} k_i+\sum_{j\in J} \ell_j,$$ and since we have assumed $J\neq\O$ we have
$$\delta_\beta< w(P')=\sum_{i=o}^\tau k_i.$$On the other hand, from $(k_0, k_1,\dots,k_\tau)\le_L (\ell_0, \ell_1,\dots,\ell_\tau)$ we get $w(P')\le\delta_\beta+(\ell_0-k_0)$. Hence, summing up, we have the following sequence of integers:
$$0\le k_0\le \gamma_1\le\dots\le\gamma_\alpha\le\delta_1\le\dots\le\delta_\beta<w(P')\le\delta_\beta+(\ell_0-k_0)\le w(Q') \le n$$
by means of which we define the mentioned profiles $P$ and $Q$.

The next figure illustrates these definitions.
\begin{figure}[htb]
\centering
\includegraphics[scale=0.61] {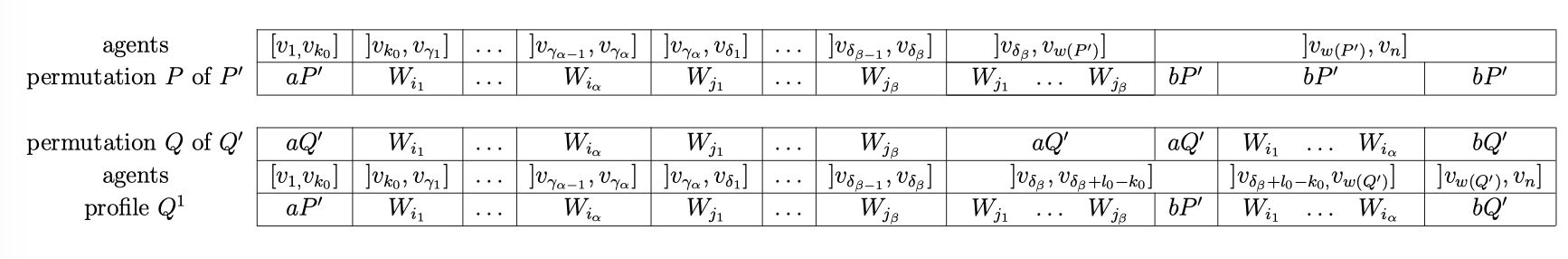}
\end{figure}

Definition of $P$:
\begin{itemize}
\item [--]on the set $[v_1, v_{k_0}]$ of agents we assign (arbitrarily) the preferences of the agents in $D(a,P')$;
\item [--]on the set $] v_{k_0}, v_{\gamma_1}]$ we assign  the preference $W_{i_1}$ to all agents;
\item [--] $\dots$
\item [--]on the set $] v_{\gamma_{\alpha-1}}, v_{\gamma_\alpha}]$ we assign  the preference $W_{i_\alpha}$ to all agents;
\item [--]on the set $] v_{\gamma_{\alpha}}, v_{\delta_1}]$ we assign  the preference $W_{j_1}$ to all agents;
\item [--] $\dots$
\item [--]on the set $] v_{\delta_{\beta-1}}, v_{\delta_\beta}]$ we assign  the preference $W_{j_\beta}$ to all agents;
\item[--] on the set of agents $] v_{\delta_\beta}, v_{w(P')}]$ we distribute the remaining preferences $W_{j_r}$ each being present $(k_{j_r}-\ell_{j_r})$ more times in the original profile $P'$;
\item[--] to the remaining agents in $]v_{w(P')}, v_n]$ we assign (arbitrarily) the preferences of the agents in $D(b,P')$.
\end{itemize}

Definition of $Q$:
\begin{itemize}
\item [--]on the set $[v_1, v_{k_0}]$ of agents we assign (arbitrarily)  preferences of $k_0$  agents in $D(a,Q')$; since preferences of remaining $(\ell_0-k_0)$ agents in $D(a, Q')$ have not been selected, we distribute them (arbirarily) to the agents of the set $]v_{\delta_\beta},v_{\delta_\beta+\ell_0-k_0}]$;
\item [--] to the agents of the sets $] v_{k_0}, v_{\gamma_1}]\dots, ] v_{\gamma_{\alpha-1}}, v_{\gamma_\alpha}], ] v_{\gamma_{\alpha}}, v_{\delta_1}],\dots, ] v_{\delta_{\beta-1}}, v_{\delta_\beta}]$ we assign  respectively the preference $W_{i_1}, \dots, W_{i_\alpha}, W_{j_1},\dots,  W_{j_\beta}$;
\item[--] on the set of agents $] v_{\delta_\beta+\ell_0-k_0}, v_{w(Q')}]$ we distribute the remaining preferences $W_{i_r}$ each being present $(\ell_{i_r}-k_{i_r})$ more times in the original profile $Q'$;
\item[--] to the remaining agents in $]v_{w(Q')}, v_n]$ we assign (arbitrarily) the preferences of the agents in $D(b,Q')$.
\end{itemize}
Let us also introduce the profile $Q^1$ identical to $P$ from agent $v_1$ till agent $v_{\delta_\beta+\ell_0-k_0}$, and identical to $Q$ on the remaining agents, i.e.\,
$Q^1=\big[P_{[v_1, v_{\delta_\beta+\ell_0-k_0}]}, Q_{]v_{\delta_\beta+\ell_0-k_0}, v_n]}\big]$

\bigskip

Now suppose that on the contrary $\phi(Q)=b$. Considering the profile $Q^1$, we have that $\phi(Q^1)$ must be $b$, otherwise the coalition $[v_1, v_{k_0}]\cup ]v_{\delta_\beta}, v_{\delta_\beta+\ell_0-k_0}]$ manipulates $Q$ by presenting $Q^1$.
If $\delta_\beta+(\ell_0-k_0)=n$, then $Q^1$ coincides with $P$ and this is not possible since $\phi(P)=a$. Therefore it must be the case that $\delta_\beta+(\ell_0-k_0)<n$, hence the coalition $]v_{\delta_\beta+(\ell_0-k_0)}, v_n]$ manipulates $P$ by presenting $Q^1$.
\eproof

 \end{document}